\documentclass[%
 reprint,
superscriptaddress,
groupedaddress,
 amsmath,amssymb,
 aps,prl
]{revtex4-2}

\usepackage{graphicx}
\usepackage{dcolumn}
\usepackage{bm}
\usepackage{amsmath, amsfonts, amsmath, amssymb}
\usepackage{mathtools}       
\usepackage{tikz}
\usetikzlibrary{
    decorations.pathmorphing, 
    decorations.markings,   
    calc,                   
    positioning             
}
\usepackage[compat=1.1.0]{tikz-feynman}

\begin{document}

\title{Feynman Tree Theorem and the Gelfand--Yaglom Formula}

\author{Ipak Fadakar, Guilherme L. Pimentel, and Behrang Tafreshi}
\affiliation{Scuola Normale Superiore and INFN, Piazza dei Cavalieri 7, 56126, Pisa, Italy}

\date{\today}

\begin{abstract}
The Gelfand--Yaglom formula equates the one-loop determinant of a Schr\"odinger operator with Dirichlet boundary conditions to the solution of an initial value problem. Following a suggestion by Polyakov, we provide a new diagrammatic proof of this formula as a direct application of the Feynman Tree Theorem. By cutting the one-loop determinant, we re-express it as a sum of tree diagrams. These trees explicitly encode the solution to the Gelfand--Yaglom initial value problem. We extend this diagrammatic framework to general boundary conditions and comment on a potential generalization to quantum field theory.
\end{abstract}

\maketitle

\textit{Introduction.}---One-loop determinants play a key role in semiclassical analysis where much of the beauty and surprise of quantum theory is already present. Important examples of such computations exist in statistical mechanics \cite{kleinert2004path}, quantum mechanics~\cite{janssen2026odeinstantons}, Quantum Field Theory (QFT) \cite{Coleman1977FalseVacuum, CallanColeman1977FalseVacuum,Dunne_2008}, and (super)gravity \cite{ColemanDeLuccia1980VacuumDecay, sen_log_correction_bh_entropy}. A non-perturbative, and in some sense classical, translation of this task was achieved by Gelfand and Yaglom \cite{GelfandYaglom1960}. The Gelfand--Yaglom formula states that the determinant of a second order ordinary differential operator $\hat{O}$ with Dirichlet boundary condition on the interval $[t_i, t_f]$, is proportional to $y(t_f)$ where 
\begin{equation}
    \hat{\mathcal{O}} y(t) = 0, \quad y(t_i)=0, \quad \dot{y}(t_i) = 1;
\end{equation}
therefore $y(t)$ is the solution of an Initial Value Problem (IVP). The ambiguity in the proportionality constant stems from the divergent nature of the functional determinant; it is conventionally resolved by normalizing against a suitable reference operator to consider a ratio of determinants. Crucially, while the determinant is constructed from the full spectral data, the Gelfand--Yaglom formula reduces it to an IVP. Intuitively, the spectral insensitivity of the IVP arises because the ultraviolet modes of $\hat{\mathcal{O}}$ and the reference operator share identical asymptotic behaviors, canceling high-energy contributions and retaining sensitivity only to low-energy spectral features. Practically, solving an IVP numerically is significantly more efficient than computing the full spectrum of a differential operator.

Standard derivations of the Gelfand–Yaglom formula follow one of two routes: a direct analytical argument \cite{Coleman:1985rnk} or a finite-dimensional construction obtained by discretizing the t-variable \cite{Dowker_2012}. Several alternative proofs, employing a variety of methods, have been developed \cite{Kleinert_1999,Ossipov_2018,Carosi_2026}, together with extensions of the formula beyond Dirichlet to more general boundary conditions \cite{Kleinert_1999,Forman1987}. Most notably, Forman \cite{Forman1987} established a far-reaching generalization which, schematically, relates ratios of functional determinants of differential operators on a manifold $\mathcal{M}$ to the Fredholm determinant of an associated pseudo-differential operator acting on boundary data on $\partial\mathcal{M}$. Crucially, because the resulting boundary operator is in general \textit{pseudo}-differential rather than purely differential, Forman's result cannot be applied recursively to further reduce the determinant to submanifolds of codimension two or higher. The Gelfand--Yaglom theorem is recovered as the one-dimensional special case, $\dim\mathcal{M}=1$, for which the boundary data are finite-dimensional. Despite these advances, whether there exists a comparably simple and computationally efficient field-theoretic analogue of the Gelfand--Yaglom formula remains an open question.

In this Letter we provide a method that potentially generalizes to quantum field theory, following a suggestion by Polyakov. We provide a diagrammatic proof that identifies the Gelfand--Yaglom formula as the quantum-mechanical counterpart of the Feynman Tree Theorem (FTT). While standard FTT applies to infinite flat space, our finite-interval formulation relies on the same core mechanism: decomposing the propagator into a retarded component and residual terms. The principal advantage of this proof is that it recasts the formula directly in the natural language of perturbative QFT: Feynman diagrams. This perspective suggests that a suitable resummation of the tree diagrams generated by applying the FTT to one-loop diagrams may provide a natural route towards a field-theoretic generalization of the Gelfand--Yaglom formula. While technical elements of our calculation share similarities with those in \cite{Gesztesy_2004}, our proof strategy and physical interpretation are fundamentally distinct. Although the extension to QFT of this derivation is straightforward, the interpretation of the resulting tree diagrams as some IVP or Cauchy problem is still lacking. Nonetheless, the proof in quantum mechanics is new and interesting, so we present it below.

\textit{Dirichlet Boundary Condition and the FTT.}---We now present our diagrammatic proof, establishing that the Gelfand--Yaglom formula is the one-dimensional, quantum-mechanical reduction of the Feynman Tree Theorem. 

Consider the perturbed harmonic oscillator on $[t_i, t_f]$ with Dirichlet boundary condition, $\hat{\mathcal{O}}_g y_g(t) = 0$, where $\hat{\mathcal{O}}_g = \partial_t^2 + \omega^2 - g u(t)$. The reference operator is chosen to be $\hat{\mathcal{O}}_0$. The exact solution to the IVP $y_g(t_i)=0, \dot{y}_g(t_i)=1$ satisfies the Volterra integral equation 
\begin{equation}
    y_g(t) = y_0(t) + g \int_{t_i}^t dt' G_R(t, t') u(t') y_g(t'),
\end{equation}
where $G_R(t, t') = \omega^{-1}\Theta(t-t')\sin\omega(t-t')$ \footnote{We take the convention $\Theta(0)=1$.}, is the retarded Green's function of $\hat{O}_0$ and
\begin{equation}
    y_0(t) = G_R(t, t_i) = \omega^{-1}\sin\omega(t-t_i),
    \label{eq: Dirichlet free solution}
\end{equation}
is the solution to the reference IVP. Iterating this equation order-by-order in $g$ generates
\begin{gather}
        y_g(t_f) = y_0(t_f) + \sum_{n=1}^\infty g^n y_g^{(n)}(t_f),\\
        y_g^{(n)}(t_f) = \int_{t_i}^{t_f} G_R(t_f, t_1) \prod_{k=1}^n  u(t_k) G_R(t_k, t_{k+1}) dt_k ,
\end{gather}
with $t_{n+1} \equiv t_i$, represented diagrammatically as a sum of tree graphs anchored at $t_f$.

Conversely, the one-loop functional determinant ratio is given by the standard perturbative expansion
\begin{equation}
    \ln \frac{\det \hat{\mathcal{O}}}{\det \hat{\mathcal{O}}_0} = -\sum_{n=1}^\infty \frac{1}{n} g^n \mathrm{Tr} \left[ (G_D u)^n \right],
    \label{eq:det_ratio}
\end{equation}
where $G_D(t, t')$ is the Dirichlet Green's function of $\hat{\mathcal{O}}_0$ on $[t_i, t_f]$. The bridge between these two representations is the exact identity connecting the Dirichlet propagator to the retarded one:
\begin{equation}
        G_D(t, t') = G_R(t, t') - \frac{G_R(t,t_i)G_R(t_f, t')}{G_R(t_f, t_i)},
    \label{eq:retarded_dirichlet}
\end{equation}
where the second term is tied to the boundary conditions; we call this term ``bulk-to-boundary" \footnote{Distinct from the holographic propagator of the same name, `bulk-to-boundary' here simply reflects that one argument of $G_R$ in the numerator is pinned to a temporal boundary ($t_i$ or $t_f$). In QFT, the analogue of this factor is an on-shell Dirac $\delta$-function.}. Because the retarded Green's function vanishes at equal times ($G_R(t, t) = 0$) and obeys causal ordering ($G_R(t_1, t_{2}) \dots G_R(t_n, t_1) = 0$), substituting Eq.~\eqref{eq:retarded_dirichlet} into the closed loop trace $\mathrm{Tr}[(G_D u)^n]$ in Eq. \eqref{eq:det_ratio}, cuts the closed $n$-point loop into linear (tree) chains. 

At each order $g^n$, expanding the product of $n$ factors of $G_D$ generates $2^n$ terms. Causal ordering eliminates closed loops, leaving only terms where at least one bulk-to-boundary propagator cuts the loop open. 

Now consider matching these cut loop diagrams against the logarithmic expansion of the IVP solution, $\ln (1 + x) = \sum_{n=1}^\infty \frac{(-1)^{n+1}}{n} x^n$, where 
\begin{equation}
\begin{aligned}
    x &= \frac{y_g(t_f)}{y_0(t_f)} - 1 = \sum_{i=1}^\infty g^i \frac{y_g^{(i)}(t_f)}{y_0(t_f)} \\[8pt]
      &=
\def\masterscale{1}
\begin{tikzpicture}[
  baseline=0pt, 
  scale=\masterscale,
  every node/.append style={transform shape}
]

\def\L{0.50}       
\def\h{0.35}       
\def\gap{0.08}     
\def\plusgap{0.18} 

\tikzset{
  fermion/.style={line width=0.5pt},
  photon/.style={
    decorate,
    decoration={snake, amplitude=1.0pt, segment length=3.2pt},
    line width=0.5pt
  }
}

\node[anchor=base west] (g1) at (0,0) {$\displaystyle g$};
\coordinate (A0) at ($(g1.east |- 0,0)+(\gap,0)$);
\coordinate (A1) at ($(A0)+(\L,0)$);
\draw[fermion] (A0) -- (A1);
\draw[photon] ($(A0)+({0.5*\L},0)$) -- ++(0,\h);

\node[anchor=base west] (p1) at ($(A1)+(\plusgap,0)$) {$\displaystyle +\; g^2$};

\coordinate (B0) at ($(p1.east |- 0,0)+(\gap,0)$);
\coordinate (B1) at ($(B0)+({1.4*\L},0)$);
\draw[fermion] (B0) -- (B1);
\draw[photon] ($(B0)+({0.35*\L},0)$) -- ++(0,\h);
\draw[photon] ($(B0)+({1.05*\L},0)$) -- ++(0,\h);

\node[anchor=base west] (p2) at ($(B1)+(\plusgap,0)$) {$\displaystyle +\; g^3$};

\coordinate (C0) at ($(p2.east |- 0,0)+(\gap,0)$);
\coordinate (C1) at ($(C0)+({1.8*\L},0)$);
\draw[fermion] (C0) -- (C1);
\draw[photon] ($(C0)+({0.3*\L},0)$) -- ++(0,\h);
\draw[photon] ($(C0)+({0.9*\L},0)$) -- ++(0,\h);
\draw[photon] ($(C0)+({1.5*\L},0)$) -- ++(0,\h);

\node[anchor=base west] at ($(C1)+(\plusgap,0)$) {$\displaystyle +\; \dots \mkern 4mu .$};

\end{tikzpicture}
\end{aligned}
\end{equation}
A given partition of $n$ vertices into $m_i$ trees of size $i$ (such that $\sum_{i=1}^n i m_i = n$) from the expansion of $\ln(1+x)$ yields the multinomial coefficient
\begin{equation}
    \mathcal{C}_{\{m_i\}} = (-1)^{\sum_{i} m_i} \, \frac{n}{\sum_{i=1}^n m_i} \binom{\sum_{i=1}^n m_i}{m_1, m_2, \dots, m_n}.
\end{equation}
Combinatorially, disregarding the sign, $\mathcal{C}_{\{m_i\}}$ counts the exact number of distinct ways to sever an unlabeled $n$-vertex closed loop into a set of $m_i$ linear tree chains of length $i$. This counting naturally incorporates the cyclic symmetry of the closed loop and the rotational equivalence of the chain of trees. Summing over all valid partitions reproduces $\mathrm{Tr}[(G_D u)^n]$ term-by-term. For example, at second order, the formula takes the following diagrammatic form:
\begin{equation}
\begin{tikzpicture}[baseline=(current bounding box.center)]

\def\R{0.38}      
\def\leg{0.38}    
\def\L{1.10}      
\def\h{0.45}      

\tikzset{
  fermion/.style={line width=0.5pt},
  photon/.style={
    decorate,
    decoration={snake, amplitude=1.1pt, segment length=3.8pt},
    line width=0.5pt
  }
}

\coordinate (C) at (0.45,0);
\draw[fermion] (C) circle (\R);
\draw[photon] ($(C)+(0,\R)$) -- ($(C)+(0,\R+\leg)$);
\draw[photon] ($(C)+(0,-\R)$) -- ($(C)+(0,-\R-\leg)$);

\node at (1.35,0) {$\displaystyle =$};

\node at (1.90,0) {$\displaystyle -2$};

\coordinate (A0) at (2.25,0);
\coordinate (A1) at ($(A0)+(\L,0)$);
\draw[fermion] (A0) -- (A1);
\draw[photon] ($(A0)+({0.42*\L},0)$) -- ++(0,\h);
\draw[photon] ($(A0)+({0.74*\L},0)$) -- ++(0,\h);

\node at (3.80,0) {$\displaystyle +$};

\coordinate (B0) at (4.25,0);
\coordinate (B1) at ($(B0)+(\L,0)$);
\draw[fermion] (B0) -- (B1);
\draw[photon] ($(B0)+({0.58*\L},0)$) -- ++(0,\h);

\node at (5.80,0) {$\displaystyle \times$};

\coordinate (D0) at (6.25,0);
\coordinate (D1) at ($(D0)+(\L,0)$);
\draw[fermion] (D0) -- (D1);
\draw[photon] ($(D0)+({0.58*\L},0)$) -- ++(0,\h);

\node at (7.55,0) {$.$};

\end{tikzpicture}
\end{equation}
Hence, resumming perturbation theory to all orders yields the exact determinant ratio, $\frac{\det \hat{\mathcal{O}_g}}{\det \hat{\mathcal{O}_0}} = \frac{y_g (t_f)}{y_0 (t_f)}$.

This completes the demonstration that the Gelfand--Yaglom formula is the exact $(0+1)$-dimensional reduction of the Feynman Tree Theorem: it reformulates a 1-loop quantum amplitude (the functional determinant) as a sum over tree-level classical paths (the IVP solution) generated by cuts. Unlike in field-theoretic applications, no phase-space or loop-integration measure appears due to the zero-dimensional nature of the boundary.

\textit{General Boundary Conditions.}---We now extend the diagrammatic proof to general linear boundary conditions. In the Dirichlet setting, the endpoint values $y_g(t_f)$ and $y_0(t_f)$ completely specify the functional determinant ratio. Moreover, the scalar factor $y_0(t)$ alone is enough to express the bulk-to-boundary term of the Dirichlet Green's function. This scalar structure is a special simplification of Dirichlet boundary data that masks the underlying two-dimensional boundary data (see Eq. \eqref{eq: general linear BC} below).

For general boundary conditions, these objects become matrix-valued, as it is more natural to express general boundary conditions for second-order equations using a pair of first-order equations. The endpoint values are replaced by $2 \times 2$ matrices encoding the boundary conditions, and the bulk-to-boundary term in the Green's function is modified. We now show that the FTT mechanism operates identically in this matrix-valued space: closed loops built from the general boundary Green's function decompose into retarded cut diagrams coupled via matrix multiplication over the $2 \times 2$ indices labeling the boundary degrees of freedom.

Consider the scalar operator $\hat{\mathcal O}_g = \partial_t^2+\omega^2-g\,u(t)$ on $[t_i, t_f]$. The scalar homogeneous equation $\hat{\mathcal O}_g y_g(t) = 0$ is written as the first-order linear system for $\mathbf y_g(t) = (y_g(t), \dot{y}_g(t))^T$:
\begin{equation}
    \partial_t \mathbf y_g(t) = A_g(t) \mathbf y_g(t), \quad A_g(t) = \begin{pmatrix} 0 & 1 \\ -\omega^2 + g u(t) & 0 \end{pmatrix}.
\end{equation}
The time-evolution operator $U_g(t,t')$ is defined by
\begin{equation}
    \partial_t U_g(t,t') = A_g(t)U_g(t,t') \quad\text{with} \quad U_g(t,t) = I,
\end{equation}
mapping configurations between any two times via $\mathbf y_g(t) = U_g(t,t') \mathbf y_g(t')$.

We impose general linear boundary conditions via $2 \times 2$ constant matrices $M_i, M_f$:
\begin{equation}
    M_i \mathbf y_g(t_i)+M_f \mathbf y_g(t_f)=0.
    \label{eq: general linear BC}
\end{equation}
Since $\mathbf y_g(t_f)= U(t_f, t_i) \mathbf y_g(t_i)$, Eq.~\eqref{eq: general linear BC} requires $\mathbb M_g \mathbf y_g(t_i)=0$, where we define the characteristic matrix of the boundary value problem:
\begin{equation}
    \mathbb M_g = M_i+M_f U_g(t_f, t_i).
    \label{eq: characteristic boundary matrix}
\end{equation}
The operator $\hat{\mathcal O}_g$ admits a zero mode precisely when $\det \mathbb M_g=0$. We assume the standard perturbative setup where the free operator $\hat{\mathcal O}_0$ has no zero modes, ensuring $\mathbb M_0$ is invertible. 

In this first-order matrix formulation, the retarded free Green's function $\mathbf{G}_R(t,t') = \Theta(t-t')\mathbf{U}_0(t,t')$ takes the explicit $2\times 2$ form \footnote{We omit an inconsequential $\delta(t-t')$ contact term.}
\begin{equation}
    \mathbf G_R = \begin{pmatrix} -\partial_{t'}G_R(t, t') & G_R(t, t') \\ -\partial_t \partial_{t'} G_R(t, t') & \partial_t G_R(t, t') \end{pmatrix}.
\end{equation} 
Furthermore, the general boundary Green's function $\mathbf G_B$ satisfies
\begin{equation}
    M_i \mathbf G_B(t_i, t') + M_f \mathbf G_B(t_f, t') = 0.
    \label{eq: general linear BC G_B}
\end{equation}
Because both $\mathbf G_B$ and $\mathbf G_R$ obey
\begin{equation}
    (\partial_t \mathbb I_2 - A_g) \mathbf G(t,t') = \delta(t-t') \mathbb I_2,
\end{equation}
they differ only by a homogeneous solution. Imposing the boundary condition \eqref{eq: general linear BC G_B} yields
\begin{equation}
    \begin{aligned}
        \mathbf{G}_B(t,t')= \mathbf{G}_R(t,t') - \mathbf{G}_R(t, t_i) \mathbb M_0^{-1} M_f \mathbf G_R(t_f, t').
    \end{aligned}
\end{equation}

Projecting back to the scalar component yields the general-boundary analogue of Eq. \eqref{eq:retarded_dirichlet}:
\begin{equation}
    \begin{aligned}
        G_B(t,t')= &G_R (t,t') \\
        &-e_1^T \mathbf G_R (t, t_i) \mathbb M_0^{-1}M_f \mathbf G_R(t_f, t') e_2,
    \end{aligned}
    \label{eq: GB equals GR minus C}
\end{equation}
where the second line acts as a bulk-to-boundary propagator (with $e_1 = (1,0)^T$ and $e_2 = (0,1)^T$).

Substituting Eq.~\eqref{eq: GB equals GR minus C} into the functional determinant expansion
\begin{equation}
    \ln(\det\mathcal O_g/\det\mathcal O_0) = -\sum_{n=1}^{\infty} \frac{1}{n}g^n \operatorname{Tr}\left[(G_Bu)^n\right],
\end{equation}
the purely retarded cycle $\prod G_R$ vanishes by causality. Every surviving term contains at least one bulk-to-boundary propagator, decomposing the loop into trees. 

To collect these contributions, the interaction potential is expressed through $\delta A_g \equiv (A_g - A_0)/g =  u(t) e_2 e_1^T$. Consequently, we define the $m$-vertex tree matrix $\mathcal X_m$ as an alternating chain of matrix propagators and interaction vertices ($t_{m+1} \equiv t_i$):
\begin{equation}
    \mathcal X_m = \mathbb M_0^{-1}M_f \int_{t_i}^{t_f} \mathbf G_R(t_f, t_1) \prod_{k=1}^m  \mathbf \delta A_g(t_k) \mathbf G_R(t_k, t_{k+1}) dt_k
    \label{eq: Dm definition}
\end{equation}
Furthermore, expanding the internal matrix products reveals that the projection $e_1^T \mathbf{G}_R(t,t') e_2$ perfectly reproduces the retarded propagators $G_R(t,t')$ between the vertices. Diagrammatically, $\mathcal X_m$ is a retarded tree carrying indices in the space of boundary degrees of freedom. The trees are then contracted via matrix multiplication. The contraction refers to the boundary indices.

Expressed in terms of the trees $\mathcal X_m$, the expansion takes the form
\begin{equation}
    \operatorname{Tr}\left[(G_Bu)^n\right] = \sum_{k=1}^{n} \frac{n}{k}(-1)^k \sum_{\substack{\sum m_j = n \\ m_j \geq 1}} \operatorname{tr}_{\partial} \left( \mathcal X_{m_1}\cdots\mathcal X_{m_k} \right),
\end{equation}
where $\operatorname{tr}_{\partial}$ denotes the $2 \times 2$ matrix trace. Resumming over $n$, matches the exact series expansion of the finite-dimensional logarithm:
\begin{equation}
    \ln\frac{\det\mathcal O_g}{\det\mathcal O_0} = \operatorname{tr}_{\partial} \ln\left[ I+ \sum_{m=1}^{\infty} g^m\mathcal X_m \right].
    \label{eq: loop equals boundary tree log}
\end{equation}
This establishes the general-boundary Feynman Tree Theorem: the 1-loop diagram is exactly the logarithm of the sum of matrix-valued trees.

To evaluate the resummed matrix, we iterate the Volterra integral equation for the time-evolution operator at $t_f$, which identifies the tree sum as
\begin{equation}
    \sum_{m=1}^{\infty} g^m\mathcal X_m = \mathbb M_0^{-1}M_f (U_g(t_f, t_i)- U_0(t_f, t_i)).
\end{equation}
Substituting this back into Eq.~\eqref{eq: loop equals boundary tree log} reduces the argument of the logarithm to $\mathbb M_0^{-1} \mathbb M_g$. Taking the matrix exponent yields the generalized Gelfand--Yaglom formula:
\begin{equation}
    \frac{\det\mathcal O_g}{\det\mathcal O_0} = \frac{\det\mathbb M_g}{\det\mathbb M_0}.
    \label{eq: generalized GY final}
\end{equation}

The standard Dirichlet result is immediately recovered upon substituting the corresponding boundary matrices $M_i$ and $M_f$.

\textit{Roadblocks to QFT.}---Although our matrix formulation is derived in one dimension, the incorporation of general boundary conditions provides a conceptual bridge to QFT. The appearance of $2\times 2$ matrices is a direct consequence of casting the \textit{second}-order differential operator into a first-order system. Presumably in a lattice regularization there should be a formulation of the Gelfand--Yaglom formula where the matrices have twice as many entries as lattice points. Moreover, the Feynman Tree Theorem can of course be applied to one-loop determinants in higher dimensions. Due to the internal phase space integrals in the resulting chains, the interpretation of the trees as solutions to an initial value problem (perhaps a Cauchy problem) remains unclear, and is under investigation.

\textit{Outlook.}---In this Letter, we have established a diagrammatic formulation connecting functional determinants directly to the Feynman Tree Theorem in one dimension. This framework provides a structural bridge between operator determinants, classical trajectories and standard perturbative techniques in quantum field theory. While extending these results to quantum field theories remains a challenge, our proof provides a roadmap forward, potentially connecting one-loop determinants to initial value problems in higher dimensional spacetimes.

\textit{Acknowledgments.}---We thank Sasha Polyakov for suggesting to use the Feynman Tree Theorem as a tool to prove the Gelfand--Yaglom formula. We are also grateful to Craig Clark, Filippo Nardi, Salvatore Raucci, Omri Rosner, Facundo Rost, Zimo Sun, and Mattia Varrone for helpful comments and discussions. IF, GLP and BT
are supported by the ERC (NOTIMEFORCOSMO, 101126304), by Scuola Normale, and by INFN
(IS GSS-Pi). GLP is also supported by the Italian Ministry of Universities and Research (MUR) under contract 20223ANFHR (PRIN2022).

\bibliography{bibtex}

\end{document}